\documentclass[aps,preprint,tightenlines]{revtex4}%
\usepackage{amsfonts}
\usepackage{amsmath}
\usepackage{amssymb}
\usepackage{graphicx}%
\newtheorem{theorem}{Theorem}

\newtheorem{conclusion}[theorem]{Conclusion}

\begin{document}
\title{Compact Gravity Wave Detector}
\author{Munawar Karim}
\email{karim@sjfc.edu}
\affiliation{Department of Physics, St. John Fisher College, Rochester, NY 14618}
\author{Kris Green}
\email{green@sjfc.edu}
\affiliation{Department of Mathematics and Computer Science, St. John Fisher College,
Rochester, NY 14618}

\begin{abstract}
An incoming gravity wave being a stress wave is a surface with intrinsic
curvature. When a light beam is parallel transported on this non-Euclidian
surface it acquires an excess phase which accumulates with each curcuit. We
calculate the separate contributions to excess phase from the wave geometry as
well as the dynamic response of mirrors in a Michelson interferometer. Using
these results and a combination of analogue and digital signal processing
techniques we show how a compact interferometer can be made sensitive to
gravity waves of amplitude density $10^{-23}/\sqrt{Hz}$ within a frequency
range $10^{-4}Hz$ to $10^{4}Hz$. As an example we describe a 10cm Michelson
interferometer designed to measure gravity waves from sources as far as the
Virgo cluster.

\end{abstract}
\date{August 14, 2003}
\maketitle

\section{}

\section{Introduction}

Following earlier attempts by Weber \cite{weber}, Forward \cite{forward} and
Weiss \cite{weiss}, several groups have been engaged in building detectors and
observatories to study gravitational radiation from astrophysical sources.
Among those which seem most likely to emit signals strong enough and often
enough to trigger current detectors are inspiralling binary neutron stars from
distances up to the Virgo cluster.

Detectors may be classified as either resonant mass or interferometric. We
will be concerned with the latter only.

Interferometric detectors rely on detecting relative phase changes between a
pair of mutually orthogonal light beams intersecting a pulse of gravitational
radiation. Interference patterns jiggle because of momentary differences in
the time-delay caused by metric perturbations due to a pulse of a passing
gravity wave.

\subsection{Sensitivity Calculations}

Interference patterns can also be disturbed by noise from Brownian motion and
radiation pressure. The gravity wave signal, to be detectable, has to overcome
these noise sources. The sensitivity is calculated by comparing contributions
to the time jitter from all sources. This method gives a direct result because
it relates to the time delay - the quantity being measured.

The interferometer arms are aligned along the $x-$ and $y-$ axes. The arms are
of length $L$ as measured in flat space. The gravitational wave described by a
four-vector $k_{\rho}=\left(  \omega,\mathbf{k}\right)  $, which is incident
along the $z-$axis, perturbs the metric by a small amplitude. The background
is a flat metric $\left(  \eta_{\mu\nu}\right)  $; the perturbed metric is:%

\begin{equation}
g_{\mu\upsilon}=\eta_{\mu\nu}+h_{\mu\nu}%
\end{equation}
A null vector represents light in the interferometer. Because of the $x-$ and
$y-$alignment of the arms we need consider only the $\left(  11\right)  $ and
$\left(  22\right)  $ components of the metric: the beam splitter is in free
fall. The mirrors may or may not be in free fall, we will treat the general case.%

\begin{align}
ds^{2}  &  =0=g_{\mu\upsilon}dx^{\mu}dx^{\nu}=\left(  \eta_{\mu\nu}+h_{\mu\nu
}\right)  dx^{\mu}dx^{\nu}\\
0  &  =-c^{2}dt^{2}+\left[  1+h_{11}\left(  \omega t-\mathbf{k\cdot x}\right)
\right]  dx^{2}%
\end{align}
The gravity wave affects both time and space components. The gravity wave does
not change the coordinate length $L$, which is altered to proper length
$L_{x}=L+\xi^{1}$. Christodoulou has shown that the wave is a stress wave and
thus has a non-linear memory \cite{chris}, this property, as well as the
measure of proper time, requires the use of proper length. Along the $x-$axis
the round-trip light travel time is:%

\begin{equation}
\tau_{rt}=\frac{2L_{x}}{c}+\frac{1}{2c}%
{\displaystyle\int\limits_{0}^{L_{x}}}
h_{11}\left(  \omega t-\mathbf{k\cdot x}\right)  dx-\frac{1}{2c}%
{\displaystyle\int\limits_{L_{x}}^{0}}
h_{11}\left(  \omega t-\mathbf{k\cdot x}\right)  dx \label{taurt}%
\end{equation}
A similar integral appears for the $y-$ axis with $h_{22}$ substituted for
$h_{11}$ and the limit $L_{y}=L+\xi^{2}$. In the transverse traceless gauge we
are working with $h=h_{11}=-h_{22}$; the arm lengths are affected in opposite
directions. $h$ is assumed to be constant over the range of the integral.

The integrals may be evaluated to obtain the difference in round-trip travel
times between the $x-$ and $y-$axes. The time difference is:%

\begin{equation}
\Delta\tau_{i}=2\frac{\xi_{i}^{1}-\xi_{i}^{2}}{c}+h\left(  t\right)
\frac{2L_{i}}{c}=\frac{\xi_{i}^{1}-\xi_{i}^{2}}{L}\tau_{i}+h\left(  t\right)
\tau_{i} \label{dtau}%
\end{equation}
where $\tau_{i}$ is the proper round-trip time for the i-th beam traversing
the interferometer:%

\begin{equation}
\tau_{i}=\left[  \frac{L_{x}+L_{y}}{c}\right]  =\frac{2L_{i}}{c}%
\end{equation}
obtained after setting $\xi_{i}^{1}=-\xi_{i}^{2}$. A general reference for
this discussion is Saulson \cite{saulson}.

In the interferometer we are describing, independent light beams traverse the
interferometer several times. Each round-trip on the wave front, since it is a
curved surface, accrues a time delay of $\Delta\tau_{i}$. Again because of
intrinsic curvature the time delay accumulates with each circuit. For $p$
round-trips the total time delay is $p\Delta\tau_{i}$ (for a constant $h$,
since $\Delta\tau_{i}\ll$ duration of the gravity wave pulse, this is a
reasonable assumption). Each circuit or sample is $\tau_{i}$ in duration. This
method of sampling requires identification of the i-th beam. The interference
intensity is recorded in discrete samples.

The metric of the incident gravity wave has intrinsic curvature (non-zero
Riemann tensor), implying a wave-front surface that is saddle-like
(fig.\ref{fig1}) (see \cite{chris}). The time delay $\Delta\tau_{i}$ may be
interpreted as the excess angle $\Delta\phi_{i}$ acquired by a\ null-vector
(light beam) parallel transported simultaneously along two closed loops ($x-$
and $y-$axes), on a surface with intrinsic curvature (fig.\ref{fig1}).

\subsubsection{Response of interferometer and mirror mount}

The interferometer responds to the incident gravity wave. The interferometer
is a rigid platform. It is accelerating upwards to counter Earth's gravity. It
is in principle not in an inertial frame. Acceleration, in principle, affects
both mirror spacing and time delay. These effects are taken care of by
equations of special relativity. For a detailed discussion and experimental
confirmation see \cite{MTW},\cite{hay}. For a system accelerating
$@g\approx10m/\sec^{2}$, times and lengths will change according to
\cite{martin}:%

\[
t=t_{0}+\left(  c/g\right)  \sinh\left(  g\tau/c\right)  \text{ and }%
x=x_{0}+\left(  c^{2}/g\right)  \left[  \cosh\left(  g\tau/c\right)
-1\right]
\]
For measurement intervals $\tau\ll1\operatorname{secs}$ there is negligible
change in either $t$ or $x$. Acceleration is not an issue, the platform is in
practice, inertial.

What is the effect of the rigid platform? The time delay Eq.(\ref{dtau}) is
the sum of contributions from time and space components.

Since the phase of a plane wave is an invariant quantity i.e., $\phi=\omega
t-\mathbf{k}\cdot\mathbf{x}=\omega^{\prime}t^{\prime}-\mathbf{k}^{\prime}%
\cdot\mathbf{x}^{\prime}$, a change in phase in general stems from a change in
both time and space components. The gravity wave has two independent effects;
(i) it alters clock rates (ii) it affects lengths. Just as the gravity wave
changes the time delay from $\tau_{i}\rightarrow\tau_{i}+h\tau_{i}$, so does
it alter the length $2L\rightarrow2L+h2L$. A quick way to see why there are
two independent sources of the phase is to follow the world lines of the light
rays emerging from the beam-splitter as they reflect off the two mirrors
(fig.\ref{fig2}). The mirrors are shown in two configurations: \ (i) in free
fall and (ii) attached rigidly to the interferometer. In the freely falling
frame of the beam-splitter the limits of the integrals in Eq.(\ref{taurt}) are
changed from $0$ to $L$ to $0$ to $L+\xi^{1}$ and $0$ to $L+\xi^{2}$ along the
$x-$ and $y-$ axes respectively. The mirrors appear displaced in the same
sense (as they must be for the ratio to be $c$, the speed of light measured by
local observers) as in the freely falling case, but less so in the case of
rigid mounts. In either case, whether in free fall or rigidly attached, there
is still a phase difference because of the intrinsic curvature in the
wave-front surface. Time and space components are affected independently. A
quantitative discussion follows.

In our calculation the beam-splitter and mirrors may be treated as test masses
in free fall, or the mirrors may be attached rigidly to the interferometer
platform. In the former case the mirrors and beam-splitter move along
individual geodesics.

When $\widetilde{h}_{jk}$ is time-varying as is the case for gravity waves,
one can calculate the response as follows: For the general case consider the
mirrors as test masses connected through a spring of coordinate length $L$,
force constant $k_{\alpha}^{\mu}$ with damping constant $b_{\alpha}^{\mu}$.
The interferometer is in free fall. We may use geodesic coordinates where the
Christoffel symbols are made zero \cite{martin} on all points on the geodesic.
The geodesic is the world line of the beam-splitter. The time coordinate is
along the instantaneous tangent to the world line of the beam-splitter. There
is an orthogonal co-moving coordinate with the mirror lying on the $x-$axis.
The quantities ($h_{\alpha}^{\mu}=h,\quad k_{\alpha}^{\mu}=k,\quad b_{\alpha
}^{\mu}=b$) have only one component . The equation of motion which describes
the mirror of mass $m$ that has an instantaneous position $x^{1}$is
\cite{weber}, with appropriate modifications:%

\begin{equation}
\frac{\partial^{2}x^{1}}{\partial t^{2}}+\frac{b}{m}\frac{\partial x^{1}}%
{dt}+\omega_{0}^{2}(x^{1}-L)=-R_{1010}^{GW}x^{1}=\frac{1}{2}\frac{\partial
^{2}\widetilde{h}_{11}}{\partial t^{2}}x^{1}=-\frac{1}{2}\omega^{2}%
x^{1}\widetilde{h}_{11} \label{x1}%
\end{equation}%
\[
\frac{\partial^{2}x^{1}}{\partial t^{2}}+\frac{b}{m}\frac{\partial x^{1}}%
{dt}+(\omega_{0}^{2}+\frac{1}{2}\omega^{2}h_{11}\sin\omega t)x^{1}=\omega
_{0}^{2}L
\]
where $\widetilde{h}_{jk}=h_{jk}\sin\omega t$ and the fundamental mode is
$\omega_{0}^{2}\equiv k/m$; $L$ is the relaxed length of the spring. This is
an inhomogeneous form of Hill's equation (for $b=0$). What we observe here is
that, under the action of the gravity wave, the mirror oscillates with a
frequency that varies sinusoidally with a frequency-dependent amplitude of
$O(h)$.

Eq.(\ref{x1}) says that the elastic properties of the mirror suspension,
whether "soft" or rigid, depending as they do on the speed of sound and
therefore on lengths and time, are modulated by\ the incident gravity wave.

We re-write this equation for $\left(  b/m\right)  \ll1$ or $Q\equiv\omega
_{0}m/b\gg1$.%

\begin{equation}
\frac{\partial^{2}x^{1}}{\partial t^{2}}+(\omega_{0}^{2}+\frac{1}{2}\omega
^{2}h_{11}\sin\omega t)x^{1}=\omega_{0}^{2}L \label{x2}%
\end{equation}
This equation may be simplified by defining a new dynamic variable
$\epsilon^{1}\equiv(x^{1}-L)/L=\xi^{1}/L$. In terms of this new variable Eq.
(\ref{x2}) becomes%
\begin{equation}
\frac{\partial^{2}\epsilon^{1}}{\partial t^{2}}+(\omega_{0}^{2}+\frac{1}%
{2}\omega^{2}h_{11}\sin\omega t)\epsilon^{1}=-\frac{1}{2}\omega^{2}h_{11}%
\sin\omega t \label{x3}%
\end{equation}

This equation describes, with appropriate choice of $\omega_{0}$, any type of
mirror suspension, whether soft or stiff. We use an iterative process to solve
this equation. Using a small expansion parameter $q=\frac{1}{2}h_{11}%
(\omega/\omega_{0})^{2}\equiv\frac{1}{2}ha^{2}$ (recall that $h_{11}\ll1$), we
assume a series solution of the form%

\begin{equation}
\epsilon^{1}(t)=%
{\displaystyle\sum\limits_{n=0}^{\infty}}
q^{n}\epsilon_{n}^{1}(t)
\end{equation}

With initial conditions $\left.  \epsilon^{1}\right\vert _{t=0}=\left.
\frac{\partial\epsilon^{1}}{\partial t}\right\vert _{t=0}=0$, to $O(h),$the
solution of Eq.(\ref{x3}) is%

\begin{equation}
\epsilon^{1}\approx\frac{1}{2}h_{11}\ \frac{\omega^{2}}{\left(  \omega_{0}%
^{2}-\omega^{2}\right)  }\ \left(  \frac{\omega}{\omega_{0}}\sin\omega
_{0}t-\sin\omega t\right)  ;\omega\neq\omega_{0} \label{tide}%
\end{equation}
For the mirror on the $y-$axis there is a similar equation:%

\begin{equation}
\frac{\partial^{2}\epsilon^{2}}{\partial t^{2}}+(\omega_{0}^{2}+\frac{1}%
{2}\omega^{2}h_{22}\sin\omega t)\epsilon^{1}=-\frac{1}{2}\omega^{2}h_{22}%
\sin\omega t
\end{equation}
with a similar solution:%

\[
\epsilon^{2}\approx\frac{1}{2}h_{22}\ \frac{\omega^{2}}{\left(  \omega_{0}%
^{2}-\omega^{2}\right)  }\ \left(  \frac{\omega}{\omega_{0}}\sin\omega
_{0}t-\sin\omega t\right)  ;\omega\neq\omega_{0}%
\]
The difference is, with $h=h_{11}=-h_{22}$:%

\begin{equation}
\epsilon^{1}-\epsilon^{2}=\frac{\xi_{i}^{1}-\xi_{i}^{2}}{L}=h\ \frac
{\omega^{2}}{\left(  \omega_{0}^{2}-\omega^{2}\right)  }\ \left(  \frac
{\omega}{\omega_{0}}\sin\omega_{0}t-\sin\omega t\right)  ;\omega\neq\omega_{0}
\label{x4}%
\end{equation}
This is a general solution which gives the time evolution of the strain for
mirrors mounted on a platform with arbitrary stiffness characterized by
resonant frequency $\omega_{0}/2\pi$. The free mass case is obtained by
setting $\omega_{0}=0$ in Eq. (\ref{x3}); the solution for which is
$\epsilon^{1}-\epsilon^{2}=h(\sin\omega t-$ $\omega t)$. The steady drift is
evident (fig.(\ref{fig3}), (different from that in ref. \cite{thorne}). A
graph of the right hand side of Eq. (\ref{x4}) is illustrative. For two
extremes of mirror suspension, soft and stiff, one can plot the time response
of the mirrors. These are shown in Figs. \ref{fig3} and \ref{fig4}. The soft
suspension emulates a free mass. The (almost) free mirrors drift away from
their equilibrium position with each passing cycle of the gravity wave,
confirming the memory effect predicted by Christodoulou \cite{chris}. By
contrast the stiff mirror undergoes miniscule periodic excursions about its
equilibrium position.

We re-write Eq. (\ref{x4}) as the excess proper time for the i-th sample:%
\begin{equation}
\frac{\xi_{i}^{1}-\xi_{i}^{2}}{L}=h\ \frac{\omega^{2}}{\left(  \omega_{0}%
^{2}-\omega^{2}\right)  }\ \left(  \frac{\omega}{\omega_{0}}\sin\omega_{0}%
\tau_{i}-\sin\omega\tau_{i}\right)  ;\omega\neq\omega_{0}%
\end{equation}
Substitution into Eq.(\ref{dtau}) gives%

\begin{equation}
\Delta\tau_{i}=\tau_{i}h+\tau_{i}h\ \frac{\omega^{2}}{\left(  \omega_{0}%
^{2}-\omega^{2}\right)  }\ \left(  \frac{\omega}{\omega_{0}}\sin\omega_{0}%
\tau_{i}-\sin\omega\tau_{i}\right)  ;\omega\neq\omega_{0} \label{dti}%
\end{equation}
for each traverse. For $p$ independent samples the total excess time is
(recall the non-Euclidian geometry of the wave surface):%

\begin{equation}%
{\displaystyle\sum\limits_{i=1}^{p}}
\Delta\tau_{i}=p\Delta\tau_{i}=p\tau_{i}h+h\ \frac{\omega^{2}}{\left(
\omega_{0}^{2}-\omega^{2}\right)  }\
{\displaystyle\sum\limits_{i=1}^{p}}
\tau_{i}\left(  \frac{\omega}{\omega_{0}}\sin\omega_{0}\tau_{i}-\sin\omega
\tau_{i}\right)  ;\omega\neq\omega_{0} \label{pti}%
\end{equation}

Time and space contributions add in phase because the speed of sound in the
platform material, sapphire is $\approx10^{4}m/\sec$; the mirrors respond
almost instantaneously i.e., within $0.10m/10^{4}m/s=10\mu\sec$, much less
than the period $1/800\sec$. The time delay can be related to the phase
difference in monochromatic light beams of wavelength $\lambda$:%

\begin{equation}
\Delta\phi_{i}=\Delta\tau_{i}\frac{2\pi c}{\lambda} \label{dphi}%
\end{equation}
The total phase difference is the sum of the time and space contributions.
Substituting into Eqs.(\ref{dtau},\ref{dphi})%

\begin{equation}
\Delta\phi_{i}=\tau_{i}h\frac{2\pi c}{\lambda}+2\frac{\xi_{i}^{1}-\xi_{i}^{2}%
}{c}\frac{2\pi c}{\lambda}=\tau_{i}h\frac{2\pi c}{\lambda}+\tau_{i}h\frac{2\pi
c}{\lambda}\frac{\omega^{2}}{\left(  \omega_{0}^{2}-\omega^{2}\right)
}\ \left(  \frac{\omega}{\omega_{0}}\sin\omega_{0}\tau_{i}-\sin\omega\tau
_{i}\right)  \ ;\omega\neq\omega_{0} \label{phii}%
\end{equation}
This is the phase-shift for each pass through the interferometer. For $p$
independent passes the accumulated phase is%

\begin{equation}
\Delta\phi=\sum_{i=1}^{p}\Delta\phi_{i}=p\Delta\phi_{i}=p\tau_{i}h\frac{2\pi
c}{\lambda}+h\frac{2\pi c}{\lambda}\left[  \frac{\omega^{2}}{\left(
\omega_{0}^{2}-\omega^{2}\right)  }\ \sum_{i=1}^{p}\tau_{i}\left(
\frac{\omega}{\omega_{0}}\sin\omega_{0}\tau_{i}-\sin\omega\tau_{i}\right)
\right]  ;\omega\neq\omega_{0} \label{pdtau}%
\end{equation}
We can evaluate the second term by writing the sum as an integral with limits
$t=0$ to $t=0.01\sec s$. Thus%
\begin{equation}
\frac{\omega^{2}}{\left(  \omega_{0}^{2}-\omega^{2}\right)  }\ \sum_{i=1}%
^{p}\tau_{i}\left(  \frac{\omega}{\omega_{0}}\sin\omega_{0}\tau_{i}-\sin
\omega\tau_{i}\right)  \rightarrow\frac{(\omega/\omega_{0})^{2}}{\left(
1-(\omega/\omega_{0})^{2}\right)  }\
{\displaystyle\int\limits_{0}^{0.01}}
d\tau\left(  \frac{\omega}{\omega_{0}}\sin\omega_{0}\tau-\sin\omega
\tau\right)
\end{equation}
For a soft suspension $\omega_{0}=2\pi$, we choose a maximum driving frequency
of $\omega=2\pi\times800$. The integral is then%
\begin{equation}
\frac{800^{2}}{\left(  1-800^{2}\right)  }\
{\displaystyle\int\limits_{0}^{0.01}}
\left(  800\times\sin(2\pi\times\tau)-\sin(2\pi\times800\tau\right)
)d\tau=\allowbreak-0.251\,25
\end{equation}
For a stiff suspension $\allowbreak\omega_{0}=2\pi\times10^{4}$ the
appropriate integral is%
\begin{equation}
\frac{0.08^{2}}{\left(  1-0.08^{2}\right)  }\
{\displaystyle\int\limits_{0}^{0.01}}
\left(  0.08\times\sin(2\pi\times10^{4}\tau)-\sin(2\pi\times800\tau\right)
)d\tau=-8.\,\allowbreak681\,9\times10^{-20}%
\end{equation}

The term in brackets in Eq. (\ref{pdtau}) is $\approx0$ for a stiff suspension
and $-0.25$ for a soft suspension. This being so we may choose a stiff mount
i.e., a high fundamental mode, which is more practical given the problems of
vibration isolation. Mirror response can be ignored.

As mirror mounts are a major obstacle in interferometric detectors, being able
to mount them in rigid platforms alleviates many problems, including noise due
to Brownian motion. For all practical purposes the total excess phase is then%
\begin{equation}
\Delta\phi=\sum_{i=1}^{p}\Delta\phi_{i}\approx p\tau_{i}h\frac{2\pi c}%
{\lambda}%
\end{equation}

We conclude from this detailed calculation that of the two independent
contributions to the excess phase- time and space components - almost the
entire contribution stems from the time component. The trajectory of the
mirrors, whether along geodesics for free mirrors, or constrained for rigid
mirrors, contributes a negligible amount to the overall phase. The excess
phase is almost entirely due to the non-Euclidian geometry of the gravity wave surface.

\subsubsection{Shot noise}

The interferometer has input and output ports. The laser power at these ports
is $P_{in}$ and $P_{out}$ respectively. The minimum sensitivity depends, among
other factors, on fluctuations of the average photon number. The average
photon flux is:%

\begin{equation}
\overline{n}=\frac{P_{out}}{\hbar\frac{2\pi c}{\lambda}}=\frac{P_{out}}%
{2\pi\hbar}\frac{\lambda}{c}\sec^{-1}%
\end{equation}
The fluctuations in the average number of photons $\overline{N}=\overline
{n}\tau$ is:%

\[
\frac{\sigma_{\overline{N}}}{\overline{N}}=\frac{1}{\sqrt{\overline{n}\tau
_{i}}}%
\]

Under operating conditions where the mean power at the output port of the
interferometer averaged over one circuit time interval is half the mean power
at the input port, i.e.,%

\begin{equation}
P_{out}=\frac{1}{2}P_{in} \label{ops}%
\end{equation}
also:%

\[
\left.  \frac{dP_{out}}{d\tau}\right\vert _{\max}=\left.  \frac{dP_{out}}%
{dL}\right\vert _{\max}\frac{dL}{d\tau}=\frac{2\pi c}{\lambda}P_{in}%
\]
each round-trip acquires a time uncertainty of:%

\[
\sigma_{\delta\tau}^{i}=\frac{\sigma_{\overline{N}}}{\overline{N}}/\frac
{1}{P_{out}}\left.  \frac{dP_{out}}{d\tau}\right\vert _{\max}=\frac{1}%
{\sqrt{\overline{n}\tau_{i}}}\frac{\lambda}{2\pi c}=\pm\sqrt{\frac
{\hslash\lambda}{P_{in}4\pi c\tau_{i}}}%
\]
For $p$ independent round-trips time uncertainties add in quadrature:%

\begin{equation}
\sigma_{\delta\tau}=\pm\sqrt{\frac{\hslash\lambda}{P_{in}4\pi c\tau_{i}}}%
\sqrt{\frac{1}{p}} \label{time}%
\end{equation}
For a shot-noise limited detector the sensitivity is obtained by requiring the
uncertainty in the shot noise to be less than the excess time delay
Eq.(\ref{pti}):%

\[
\sigma_{\delta\tau}=\sqrt{\frac{\hslash\lambda}{P_{in}4\pi cp\tau_{i}}}\leq
p\tau_{i}h
\]
The equivalent minimum detectable metric perturbation for a laser with
wavelength $\lambda$ is:%

\[
h\geq\sqrt{\frac{\hslash\lambda}{P_{in}4\pi c}}\sqrt{\frac{1}{p\tau_{i}}}%
\frac{1}{p\tau_{i}}%
\]
This expression assumes that shot noise is dominant; Brownian noise and
radiation pressure noise are negligible; generally true for the example under consideration.

\section{Detection Scheme}

\subsection{Over-sampled detector}

The expected signal is a pulse of duration $\tau\sim10ms$ of $h\sim10^{-21}$
centered at a frequency of $\sim800Hz$.

We can choose a sampling frequency. Generally, the minimum sampling frequency,
called the Nyquist frequency, is twice the bandwidth of the signal to be
sampled. For example, a signal with a bandwidth of 100Hz need be sampled every
200Hz, under ideal conditions, to be reproduced flawlessly. Practical
considerations dictate the technique of "over-sampling", that is, sampling at
rates which are integer multiples of the Nyquist frequency. The original
signal is reproduced from sampled segments. Sampling permits extraction of
signals even when the light spends one or more gravitational wave periods in
the interferometer.

Light from a laser enters a Michelson interferometer. It passes through a
beam-splitter, reflects off two mirrors onto a photodiode where the
interference intensity is recorded (fig.\ref{fig5}); the beam exits the
interferometer. The averaged intensity is that of the i-th sample. The
photodiode output is averaged over an interval $\tau_{i}$. Each sample is
independent. $\tau_{i}$ is the round-trip travel time for the i-th beam.

The over-sampling frequency is $f_{S}\sim GHz$ for the example we will
describe, compared with the signal bandwidth $f_{B}=800Hz$. Thus the
over-sampling frequency is $10^{6}$ times the Nyquist frequency.

We give an example of what can be achieved in an $L_{i}=10cm$ Michelson
interferometer. Using a $1$ watt light source of wavelength $0.545\mu m$, and
sampling every $\tau_{i}=\left(  2L_{i}/c\right)  =(2/3)\times10^{-9}%
\operatorname{secs}$.

Since the duration of the pulse is expected to be $10^{-2}\operatorname{secs}%
$, the product $p\tau_{i}=10^{-2}\operatorname{secs}$. One may collect as many
as $1.5\times10^{7}$ discrete samples if we choose to integrate over the
entire length of the pulse. The sensitivity is:%

\begin{equation}
h\geq\sqrt{\frac{\hslash\lambda}{P_{in}4\pi c}}\sqrt{\frac{1}{p\tau_{i}}}%
\frac{1}{p\tau_{i}};\omega\neq\omega_{0}%
\end{equation}%
\begin{equation}
h\geq10^{-22} \label{sens}%
\end{equation}
a result which depends only on the pulse duration but is noticeably
independent of the sampling interval $\tau_{i}$, and necessarily, the length
$L_{i}$. This is also the result expected, since it confirms that the excess
phase at the end of one long circuit is equivalent to the phase accumulated in
many smaller circuits. The sensitivity is determined by the non-Euclidian
geometry of the gravity wave surface alone.

The equivalent length of the interferometer is \textit{1500km}, half the
distance a light beam travels during the $10m\sec$ gravitational wave pulse;
this is also the optimum length. The sensitivity is sufficient to detect
putative events from distances up to the Virgo cluster.

Unlike a Fabry-Perot cavity or a Herriot delay line where the light beam is
recorded after it undergoes multiple reflections in a cavity, we note that the
light beam is averaged, and recorded at the end of \textit{each round-trip}.
The sampled data stream is processed in accordance with algorithms used for
over-sampled detection. The proposed design differs in this aspect from folded interferometers.

\subsection{Proposed Design}

Once we realize that the sensitivity is independent of $L$, the interferometer
arms are chosen for convenience to be $10cm$ long; the design options also
allow some flexibility. For one; a small interferometer is easier to build,
easier to control the environment (temperature, pressure, isolation from
external noise etc.). The entire interferometer can be mounted on single
platform. Sampling the signal at the end of every round-trip prevents the
accumulation of excess phase by using the beams only once.

Mechanical stability is facilitated by mounting all the components on a $15cm$
diameter, $2cm$ thick sapphire disk. Sapphire is a suitable material because
of its low thermal expansion coefficient ($\alpha\sim10^{-6}/C^{\circ}$),
excellent thermal conductivity ($0.4watts/cmK^{\circ}$), low mechanical
dissipation ($Q\approx10^{9}$, reduces Brownian noise), stiff Young's modulus
($\sim$GPa) and a high speed of sound ($10^{4}m/s$) \cite{sapphire}, so a high
fundamental vibration frequency which facilitates isolation from external vibrations.

\subsubsection{Noise from Brownian motion and radiation pressure}

We can estimate the Brownian noise contribution. We use the
fluctuation-dissipation theorem. The mirrors, which need be no more than $5mm$
in diameter, in order to maintain a high fundamental frequency, may be
sculpted directly into the sapphire disk. The entire platform vibrates due to
thermal excitation.

The power spectrum of the fluctuation force is%
\begin{equation}
F_{th}^{2}=4k_{B}T\operatorname{Re}(Z) \label{Fl}%
\end{equation}
in terms of the impedance $Z$. The appropriate equation of motion of the
mirror is%
\begin{equation}
m\frac{\partial^{2}\xi}{\partial t^{2}}+b\frac{\partial\xi}{\partial t}%
+k\xi=F_{ext}%
\end{equation}
Or in terms of instantaneous velocity $v=i\omega\xi$:%
\begin{equation}
i\omega mv+bv-\frac{ikv}{\omega}\equiv Zv=F_{ext};\qquad\therefore Z=i\omega
m+b-\frac{ik}{\omega}%
\end{equation}
Squaring and substituting into Eq.(\ref{Fl}) we get%
\begin{equation}
\xi_{th}^{2}=\frac{4k_{B}T}{\omega^{2}}\operatorname{Re}\left(  \frac{1}%
{Z}\right)  =\frac{4k_{B}T}{\omega^{2}}\frac{b}{b^{2}+\left(  m\omega-\frac
{k}{\omega}\right)  ^{2}}%
\end{equation}
Equating $b=\omega_{0}m/Q$ and $k=m\omega_{0}^{2}$, the simplified expression
becomes%
\begin{equation}
\xi_{th}^{2}=\frac{4k_{B}T}{\omega^{2}}\frac{\omega_{0}}{mQ}\frac{1}{\left(
\frac{\omega_{0}}{Q}\right)  ^{2}+\left(  \omega-\frac{\omega_{0}^{2}}{\omega
}\right)  ^{2}}\approx\frac{4k_{B}T}{\omega^{2}}\frac{\omega_{0}}{mQ}%
\frac{\omega^{2}}{\omega_{0}^{4}}%
\end{equation}
for the chosen values of $Q$ and $\omega\ll\omega_{0}$. Thus far below the
fundamental resonance $(\omega\ll\omega_{0})$ the mirror amplitude excursion is:%

\begin{equation}
\xi_{rms}^{B}=\sqrt{\frac{4k_{B}T}{Qm\omega_{0}^{3}}}=\pm8.17\times
10^{-21}m/\sqrt{Hz}%
\end{equation}
for $T=300K$, $Q\sim10^{6},$ $m=1kg,$ $\omega_{0}\approx2\pi\times10^{4}$.
After $p$ discrete samples each of interval $\tau_{i}$, the average excursion is%

\begin{equation}
\xi_{rms}^{B}=\sqrt{\frac{4k_{B}T}{Qm\omega_{0}^{3}}}\sqrt{\frac{1}{\tau_{i}%
p}}=\sqrt{\frac{4k_{B}TQ^{-1}}{\omega m\omega_{0}^{2}}}\sqrt{\frac{c}{2L_{i}}%
}\sqrt{\frac{2L_{i}}{10^{-2}c}}=\pm8.17\times10^{-20}m
\end{equation}
$\xi_{rms}^{B}$ depends on the total sampling interval, independent of the
number of samples or the individual sampling interval. The time-jitter due to
Brownian motion is $\pm\xi_{rms}^{B}/c$:%

\[
\xi_{rms}^{B}/c=\pm2.72\times10^{-28}\operatorname{secs}%
\]
For the example under consideration the shot noise time-jitter is
Eq.(\ref{time})%

\[
\sigma_{\delta\tau}=\pm\sqrt{\frac{\hslash\lambda}{P_{in}4\pi c\tau_{i}}}%
\sqrt{\frac{1}{p}}=\pm10^{-24}\operatorname{secs}%
\]
The detector sensitivity Eq.(\ref{sens}) is limited by shot-noise in the range
of frequencies $10^{-4}Hz<\omega/2\pi<10^{4}Hz.$The upper frequency limit is
set by the fundamental mode of the platform and the lower limit by tidal frequencies.

Radiation pressure will also inject random vibrations in the mirrors. The
fluctuating power spectrum in this case is%
\begin{equation}
F_{R}^{2}=2\pi\hslash\frac{c}{\lambda}\frac{P_{in}}{c^{2}}=2\pi\hslash\frac
{c}{\lambda}\operatorname{Re}(Z)
\end{equation}
The amplitude excursion, for $p$ discrete samples each of duration $\tau_{i}$,
appropriate for $\omega\ll\omega_{0}$ is:%

\begin{equation}
\xi_{rms}^{R}=\sqrt{\frac{2\pi\hslash c}{\lambda mQ\omega_{0}^{3}}}\sqrt
{\frac{1}{\tau_{i}}}\sqrt{\frac{1}{p}}=\pm3.9\times10^{-19}m \label{rad}%
\end{equation}
The time uncertainty is:%

\[
\xi_{rms}^{R}/c=\pm10^{-27}\operatorname{secs}%
\]
Again this is negligible compared with the time uncertainty due to shot noise.
Noise from Brownian motion and radiation pressure can be ignored.

It is worth repeating that the light beams reflect off each mirror only once,
then they strike the photo-diode and are removed from the interferometer.
Light from single reflections is corrupted by fluctuations in the mirror
position due to radiation pressure and/or shot noise. Because each reflection
is independent, the noise contributions are also independent. They add in
quadrature. By contrast the situation in folded interferometers which utilize
multiple reflections, is different because mirror noise fluctuations
accumulate with each reflection, in practice limiting the number of folded
beams \cite{saulson}.

\subsection{Measurement Method}

The photodiode detects interference fringes, actually a circular region which
may be all shades from completely dark to completely bright. The photodiode
output is sampled every nanosecond. To make data handling manageable, the
photodiode output is fed first into an analogue integrator with a time
constant which is a small fraction of the expected (reciprocal) signal
frequency. For example a $10\mu\sec$ time constant is sufficient for 125
samples of an $800Hz$ signal or 1000 samples of a $100Hz$ signal. Following
the integrator the averaged signal is sent to a 16-bit analogue to digital
converter and from there onto a library for comparison against different
signal templates.

Referring to the earlier mention of a mode of operation Eq.(\ref{ops}), it
turns out that refinements are needed to operate the interferometer as a null
detector. Pockels cells or air columns, need to be inserted to provide phase
modulation. The interferometer operation point is moved to a dark fringe
\cite{weiss}. Because of the stiff mounting of the mirrors, it may be possible
to maintain the dark fringe condition by further slow modulation of the
Pockels cells or the pressure in the air columns..

The small size of the interferometer (diameter 15cm.) facilitates operation in
a vacuum environment ($10^{-7}$T is sufficient for $h\approx10^{-22}$); it
also opens up several schemes to isolate it from seismic vibrations which is a
major source of noise at low frequencies. Isolation from ground vibrations is
made easier because the design fundamental mode is $\omega_{0}=2\pi
\times10^{4}$ as is reduction in Brownian noise. This is one great advantage
in using rigidly mounted mirrors and a high-Q platform.

The interferometer needs a laser source stabilized against both intensity and
frequency drifts. Use of sapphire also minimizes platform distortions due to
temperature inhomogeneities. Instruments will be needed to monitor and control
the temperature.

The scheme described here can be extended to a three-axis detector, which may
be replicated and installed as an antenna array on several optimal locations
on Earth. An array opens up the possibility of correlation interferometry
which is a means to study the quantum properties of gravitational radiation.

\begin{conclusion}
Working with the non-Euclidian geometry of incoming gravity waves and, using
signal processing techniques we have designed a table-top 10cm Michelson
interferometer with sufficient sensitivity to detect gravitational waves from
infalling binary neutron stars from as far as the Virgo cluster.

\begin{acknowledgments}
We acknowledge with gratitude an anonymous referee for planting the seed of an
idea, and M. Bocko for technical advice.
\end{acknowledgments}
\end{conclusion}

%

\begin{figure}
[ptbh]
\begin{center}
\includegraphics[
natheight=3.037200in,
natwidth=4.498700in,
height=3.0813in,
width=4.5515in
]%
{../WINDOWS/Desktop/FFdrawings/wavesurf.ai}%
\caption{Surface of gravity wave-front. Shown is the field amplitude
$h_{\mu\nu}$ along the z-axis$:$ at this instant $h_{11}$ is negative and
$h_{22}$ is positive. The surface has intrinsic curvature. An excess phase
appears in a null vector parallel transported in a circuit on this surface.
$\Delta\phi_{i}$ is the difference in excess phase between a circuit along the
$x-$ and $y-$ axes.}%
\label{fig1}%
\end{center}
\end{figure}
%

\begin{figure}
[ptb]
\begin{center}
\includegraphics[
natheight=7.772900in,
natwidth=9.735200in,
height=3.845in,
width=4.8084in
]%
{../WINDOWS/Desktop/FFdrawings/world-lines.ai9.ai}%
\caption{World lines of rays and interferometer components. Two mirror mounts,
free in light and rigid in heavy lines, are shown under the influence of a
gravity wave of constant amplitude $h_{\mu\nu}$. Although the surfaces are
shown as planes they are saddle-shaped as in figure 1. The reflection points
depend on the type of mirror mount, and so does the excess phase.}%
\label{fig2}%
\end{center}
\end{figure}
%

\begin{figure}
[ptbh]
\begin{center}
\fbox{\includegraphics[
natheight=2.000300in,
natwidth=3.000000in,
height=2.0003in,
width=3in
]%
{HJIFJW02.wmf}%
}\caption{Mirror response in units of gravitational wave amplitude $h$ to
incoming gravity wave of maximum frequency $\omega=2\pi\times800s^{-1}$ and
duration $10ms$. Suspension frequency is set at $1Hz$.}%
\label{fig3}%
\end{center}
\end{figure}
%

\begin{figure}
[ptbh]
\begin{center}
\fbox{\includegraphics[
natheight=2.000300in,
natwidth=3.000000in,
height=2.0003in,
width=3in
]%
{HJIFLG03.wmf}%
}\caption{Mirror response in units of $h$ to incoming gravity wave of maximum
frequency $\omega=2\pi\times800s^{-1}$ and duration $10ms$. Suspension
frequency is set at $10^{4}Hz$.}%
\label{fig4}%
\end{center}
\end{figure}

\bigskip

\bigskip\pagebreak

\bigskip

\bigskip%

\begin{figure}
[ptb]
\begin{center}
\includegraphics[
natheight=6.145400in,
natwidth=10.141700in,
height=3.845in,
width=6.3261in
]%
{../WINDOWS/Desktop/FFdrawings/SGWI.ai9.ai}%
\caption{Schematic arrangement of interferometer components. Coherent beams
reflect off the $x-$ and $y-$ mirrors; when they strike the photodiode they
exit the interferometer. Output is sampled in discrete intervals: the
intensity pattern is reconstructed using sampling algorithms. Each sample is
independent, there is no multiple reflection.}%
\label{fig5}%
\end{center}
\end{figure}

\bigskip
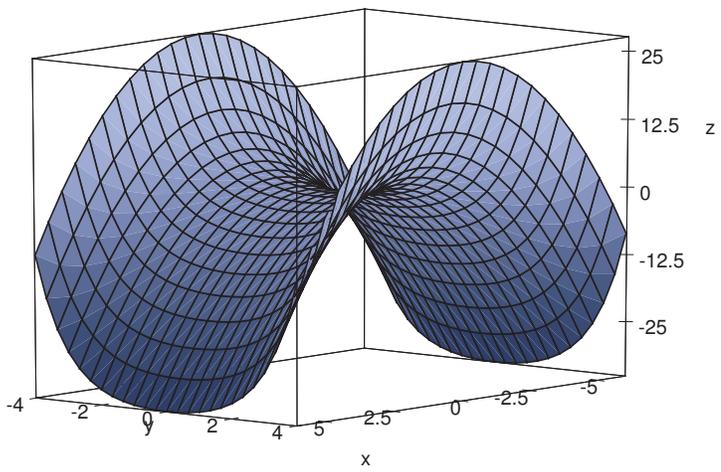
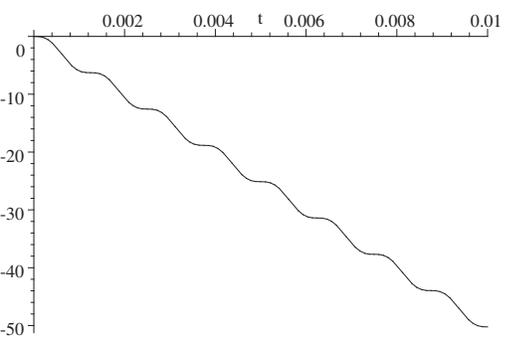
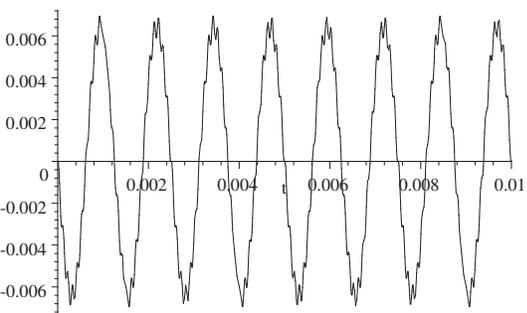
\end{document}